\documentclass[fleqn,usenatbib]{mnras}
\usepackage{graphicx}
\usepackage{amsmath}
\usepackage{natbib}
\usepackage{physics}
\usepackage{color}

\newcommand{\Sec}[1]{\S\ref{#1}}

\newcommand{\Fig}[1]{Figure~\ref{#1}}
\newcommand{\Tab}[1]{Table~\ref{#1}}

\def\bl{Babcock--Leighton }

\def\we{Waldmeier effect}

\usepackage{newtxtext,newtxmath}

\usepackage[T1]{fontenc}

\title[Prediction of Solar Cycle 25]{Physical link of the polar field build-up with the Waldmeier effect broadens the scope of early solar cycle prediction: Cycle 25 is likely to be slightly stronger than Cycle 24}

\author[Kumar et al.]{
Pawan Kumar,
Akash Biswas,
Bidya Binay Karak\thanks{E-mail: karak.phy@iitbhu.ac.in}\\
Department of Physics, Indian Institute of Technology (Banaras Hindu University), Varanasi 221005, India
}

\date{Accepted XXX. Received YYY; in original form ZZZ}

\pubyear{2022}

\begin{document}
\label{firstpage}
\pagerange{\pageref{firstpage}--\pageref{lastpage}}
\maketitle

\begin{abstract}
Prediction of the solar cycle is challenging but essential because it drives space weather. Several predictions with varying amplitudes of the ongoing Cycle~25 have been made.
We show that an aspect of the Waldmeier effect (WE2), i.e., a strong positive correlation between the rise rate and the amplitude of the cycle, has a physical link with the build-up of the previous cycle's polar field
after its reversal.
We find that the rise rate of the polar field is highly correlated with the rise rate and the amplitude of the next solar cycle. Thus, the prediction of the amplitude of the solar cycle can be made just a few years after the reversal of the previous cycle's polar field, thereby extending the scope of the solar cycle prediction to much earlier than the usual time. Our prediction of Cycle 25 based on the rise rate of the previous polar field is $137\pm 23$, which is quite close to the prediction $138\pm 26$ based on the WE2 computed from the available 2 years sunspot data of the ongoing cycle.
\end{abstract}

\begin{keywords}
Sun -- activity -- magnetic fields -- sunspots -- dynamo
\end{keywords}



\section{Introduction}
The dynamic solar magnetic field is responsible for producing energetic events like solar flares and coronal mass ejections.
These events drive space weather which sometimes has hazardous impacts on 
our space-based society.
In a strong cycle, we observe more such events and thus large impacts on the space weather. Hence, predicting the solar cycle strength is of our utmost importance. 

As the solar cycle is irregular, the prediction is challenging.
Several methods have been applied to predict the amplitudes of the past few cycles and it is not an exception for the Cycle 25 \citep{Petrovay20, Nandy21}. Out of these methods, precursor, in which the information of the previous cycle is used to predict the strength of the cycle, is the most widely used method; see \citet{Hat02, CS07, Kane10, Haz15} and Section 2 of \citet{Petrovay20}.

One important feature of the solar cycle is the \we\ \citep{W35}, which says that strong cycles take less time to rise, and vice versa. While this correlation is somewhat poor and even difficult to establish \citep[][hereafter KC11]{Dik08, KC11}, there exists a robust correlation between 
the rise rate (slope) and the amplitude of the cycle \citep{CS08}.
This correlation exists strongly in all the observed proxies of the solar cycle. KC11 called these two correlations, i.e., the correlations between the rise time and the rise rate of the cycle with the amplitude as WE1 and WE2, respectively. We mention that the \we\ is not even limited to our sun only, some other sun-like stars do show this feature \citep{garg19}.

As the rise rate can be computed when the solar cycle has just passed the minimum by a few years, we can apply WE2 to predict the amplitude of the solar cycle when the cycle is still growing and has not reached its peak. The current Cycle 25 has passed about 2~years and thus we can predict the amplitude of Cycle 25. This is one of the motivations of the present Letter. 

While WE2 is derived based on the observed correlation, there is a strong physical basis for this. KC11 showed that WE2 was robustly reproduced in the \bl\ type flux transport dynamo models with stochastic fluctuations in the poloidal source. Observations, as well as the dynamo models, suggest that if the polar field at the solar minimum is strong, then the amplitude of the next cycle will be strong \citep{MMS89, JCC07, WS09, KO11, Muno13, Priy14, KM17, KMB18, KKV21}.   
On the other hand, if a cycle is strong, then it rises fast (WE2). Hence, there is a link between the polar field at the solar minimum and the rise rate of the next cycle. We shall explicitly demonstrate this link in the present study.

The most interesting feature that we have found is that the rise rate of the polar field build-up (after its reversal) also determines the rise rate of the next sunspot cycle and thus the amplitude of the cycle.
Hence, we do not even need to wait for the time of the solar cycle minimum or the time of the peak of the polar field \citep[which is the usual time for the prediction;][]{Sch78, CCJ07} to get an idea of the next cycle strength, the rate at which the polar field develops carry this information. 
In this Letter, we shall present this link both from the observed and the dynamo model data and discuss the physical reason based on the \bl\ dynamo.
Finally, we shall predict the amplitude of the ongoing solar Cycle 25, separately using the rise rate of the current solar cycle and the rise rate of the previous cycle's polar field. We shall show that the prediction made from these two methods are very close to each other because the physics behind these two are linked.

\section{Data and Methods}
\label{sec:method}
For our analysis, we have used the monthly sunspot number (SSN) and sunspot area (SSA) data. The SSA data however are not available for Cycle 25.
For the observational measures of solar polar field, we have included the polar field strength data (monthly binned) collected from Wilcox Solar Observatory (WSO).
To remove the high fluctuations in the data of SSA and SSN, we have used Gaussian smoothing filter with FWHM = 13 and 7 months, respectively \citep{Hat02}.
As the current solar cycle 25 has only undergone 2 years from its minimum, we can calculate the rise rates based on this two years data. Hence, to make it uniform for all 13 cycles (Cycle:12--24), 
we have computed the rise rate for first 2 years of their rise phase only. As the data are not very smooth, 
and there is some overlap between two consecutive cycles during the first few months of a cycle \citep{CS08}, we excluded the first six month's data from our analysis to avoid any contamination in the rise rates due to these reasons. Further, as the rise rate of a cycle is dynamic throughout its evolution, we computed the rates at different phases with different time intervals 
(6 to 18 months, 12 to 24 months and 6 to 24 months) and finally we average these values to get one rise rate for each cycle.
For computing the rise rate of the polar field, we compute it within the first three years after the reversal, as there is no overlap between two consecutive cycles in the polar field data. 
Again here also we compute the rates at  different intervals (0 to 36 months and 12 to 36 months for north; 0 to 36 months and 24 to 36 months for south) and average those to get one rise rate for a cycle.

%
\section{Results and Discussion}
\subsection{WE2 and the prediction of Cycle 25}
\label{sec:pred_we2}
\begin{figure}
\centering
\includegraphics[scale=0.3]{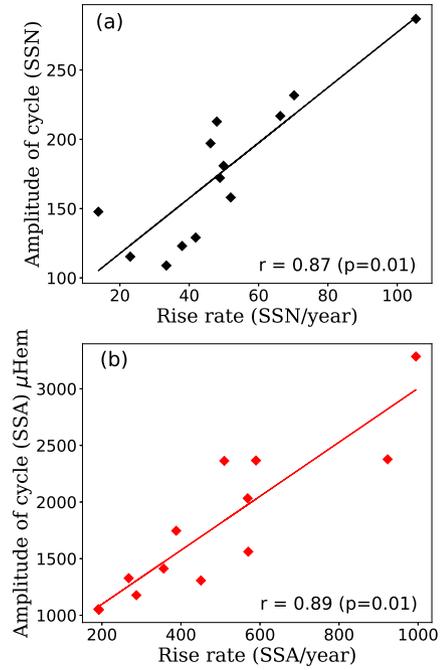}
\caption{Scatter plots between the rise rates and the amplitudes of the cycles for SSN (a) and SSA (b).
Lines are the linear regressions: $ y = m x  + c$, where 
$m = 2.001 \pm 0.308$ and $c = 77.305 \pm 16.545$ for SSN and  $m = 2.423 \pm 0.114$ and $c = 596.789 \pm 62.028$ for SSA.
}
\label{fig:corrplot}
\end{figure}

 \Fig{fig:corrplot} shows the scatter plots of the rise rates with the amplitudes of the SSN (a) and SSA (b).
 We find a strong correlation between these two quantities for both the data with linear (Pearson) correlation coefficients of 0.87 for SSN and 0.89 for SSA data. These results reproduce WE2 \citep{W35, KC11}. The straight lines in \Fig{fig:corrplot} are obtained from the linear
 regression based on Bayesian probabilistic approach (using Python’s Pymc3 routine); see figure caption for the fitted parameters. 
 The strong correlation between these quantities in \Fig{fig:corrplot} implies that if the rise rate of a cycle is known even for some part of its rising phase, then the amplitude of the cycle can be predicted well in advance. 
 To test the reliability of this prediction method, we predict the amplitudes of the last few observed cycles and compare them with observed values.
 We note that when we predict the amplitude of a given cycle, we exclude the data for that cycle while computing the regression relation. In  \Tab{table1}, we mention our predicted peak values along with their errors for previous 6 cycles (Cycle: 19--24) and the actual observed values.   
 We can see that the the predicted values are not too far from the actual ones. For some cycles, like Cycle 22, the predicted value is  quite different from the observed ones, but considering the error in the regression, it is not too much off from the allowed range.

 We do the same exercise using SSA data and the predicted amplitude of SSA are given in \Tab{table1}.
 However, in this case, we see a somewhat larger deviation in predicted values from the actual observations, although the correlation between the rise rate and the observed amplitude is better than that in the SSN data. 
 To compare these predicted peak areas 
 with the observed SSN, we 
 convert the predicted SSA into the SSN by employing the regression relation 
 (${\rm SSN} = 0.076 {\rm SSA} + 39.717$) between 
 SSN and SSA, which are  
 listed in the last column of \Tab{table1}.
 Overall, prediction based on the rise rate of both sunspot number and area supports our idea.

\begin{table}
\centering
\caption{Predictions of the solar cycle {\it amplitudes} using the rise rates of SSN and SSA (in $\mu$Hem) for the last few known cycles and the ongoing Cycle~25.
}

\begin{tabular}{lllcclcccl} 
\cline{1-6}
Cyc. & Obs. & Predicted & Obs.  & Predicted & SSN from \\
No &  SSN & peak SSN & SSA  & SSA & Col. 5   \\ 

\cline{1-6}
 25 &  $---$ &  $138\pm 26$ & $---$ & $---$ & $---$ \\
\cline{1-6}
 24 &  116   & $125\pm 26$ &   1054.0    & $1058.2\pm 84.6$ &  $121\pm 17$  \\
    
\cline{1-6}

23 & 181   & $177\pm 33$ & 1746.1 & $1536.3\pm 83.1$ & $157\pm 17$ \\
\cline{1-6}
22  &  213  & $170\pm 30$ & 2354.5	&	$2026.2\pm 129.3$	&	$195\pm 20$ \\
\cline{1-6}
21	&	   232 & $216\pm 40$ & 2363.1	&	$1831.5\pm 122.3$ 	 &     $180\pm 19$  \\
\cline{1-6}
20	&	   158 & $183\pm 32$ &  1561.4 &	$1978.8\pm 128.8$ 	&	$191\pm 20$ \\
\cline{1-6}
19	&  287	& $290\pm 72$ &  3285.2	 &     $3008.6\pm 177.2$	  &  	$270\pm 24$ \\
\cline{1-6}

\end{tabular}
\label{table1}
\end{table}

\begin{figure*}
\includegraphics[scale=0.3]{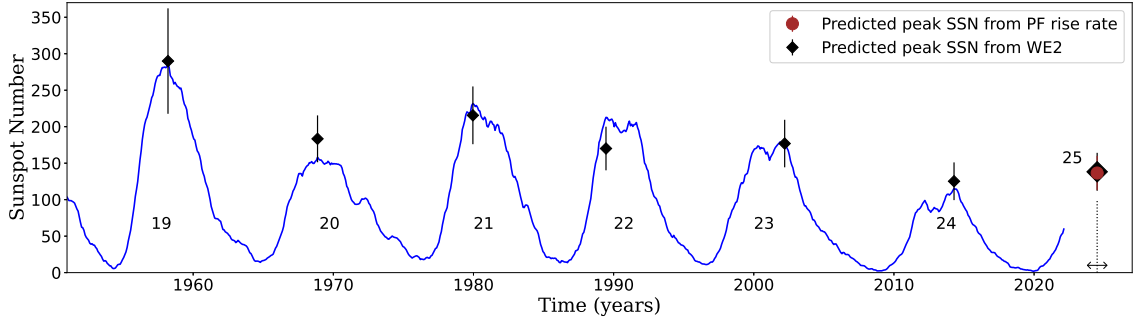}
\caption{
{
Comparison of our prediction with observations. 
Temporal variation of the observed SSN is shown by the blue curve.
The predicted amplitudes are shown by black squares and their errors by vertical lines. The time of the peak of the predicted Cycle 25 is shown by the vertical dotted line with the error by a horizontal arrow. 
The prediction for Cycle~25 using {\it the rise rate} of the previous cycle's polar field is shown by a (dark red) filled circle.
}
}
\label{fig:cycles}
\end{figure*}
\begin{figure}
\centering
\includegraphics[scale=0.3]{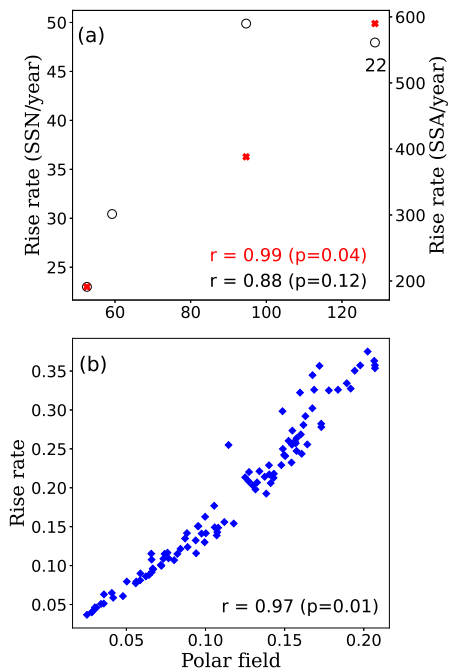}
\caption{Scatter plots between the polar field at solar cycle minimum and the rise rate of next cycle. (a) Open circles and red crosses represent the rise rates  calculated from SSN and SSA, respectively. (b) Same as (a) but from a \bl\ type dynamo model.}
\label{fig:pol}
\end{figure}

Finally we predict the peak of the ongoing Cycle 25 based on its available SSN data.
We find the predicted amplitude of Cycle 25 to be $138\pm26$; see \Tab{table1}. 
As we do not have SSA data for this cycle, we cannot predict the peak of SSA for Cycle 25 directly from the rise rate of SSA data.

 WE2 relation also gives how much time the cycle will take to reach its peak from its minimum. 
 For Cycle 25, this value comes to be $4.5\pm 0.8$ years, which is quite close to the 
 average time of $4.58\pm 0.81$ years between the cycle minimum to maximum 
 as reported in \citet[][see their Table~1]{pawan21}. 
 Therefore, we predict that the Cycle~25 will attain peak at $2024.5\pm 0.8$.
 For better visualization, our predicted peak SSNs with their error ranges are shown 
 in \Fig{fig:cycles}.

\subsection{Connecting WE2 with the previous cycle polar field}
We would like to mention that although our prediction in \Sec{sec:pred_we2} is based on an empirical relation which holds `statistically' and hence the prediction for certain  cycles (like cycle 22) may not perfectly agree with the observation. However, we still make prediction because our method is based on a strong physical ground.
 It was shown that WE2 relation (on which our prediction is based) is a robust feature of solar dynamo \citep{KC11, pipin11, pipin11b}.
Particularly, KC11 found a strong correlation for WE2 in all the simulations they have performed. They explained that fluctuations in the generation of poloidal field (\bl\ process), makes the polar field at the solar minimum unequal for different cycles. As the polar field gives rise to the toroidal field and the sunspot for the next cycle, strong polar field makes the next cycle strong. This is also established in the observational data \citep{MMS89,CCJ07, JCC07, WS09, KO11, Muno13, Priy14}.
Finally if the cycle is strong, it has to rise fast.

Based on above discussion, we expect that the rise rate of a cycle should be directly linked with the poloidal field at the cycle minimum. To check this link, we compute the linear correlation between these two, based on the observed polar field data for last four cycles; see \Fig{fig:pol}(a). We observe a reasonably good correlation between these two quantities from SSN and SSA 
data. However, the data of SSN for Cycle 22 is showing some deviation from the linear trend. This 
 may be due to the fact that we are calculating the rise rate based on only first two years data (to make it consistent with the available data for Cycle 25).
Furthermore, the polar field data are not always perfect due to limited observations in the polar regions \citep{Bertello14, Mord22}. Hence, the reliability of this relation cannot be endorsed with limited data.


Therefore, we try to explore this link between the polar field and the rise rate of the next cycle using \bl\ type dynamo theory.
To do so, we have taken the 
data from the dynamo model: Run 2DR2 as presented in \citep{pawan21} which is produced using {\it{ Surya}} code \citep{CNC04}.
We compute the correlation between the polar field at a cycle minimum and the rise rate of the following cycle from the data of 100 cycles in the same manner as done for the observed data. We find a high correlation as seen in \
\Fig{fig:pol}(b). This supports the fact that a strong poloidal field indeed makes the following cycle rise faster and hence the cycle becomes stronger, obeying WE2.
So, we believe that our prediction for the ongoing cycle 25 should be comparable
with the prediction made by other groups using the observed polar field data
but not the same because there is one more physical process involved in it that we shall discuss in \Sec{sec:pol_rr}.
To facilitate the comparison, we enlist our predicted values of the amplitude and time of the peak for cycle 25 with various other groups in \Tab{table2}.
We find that our value is slightly larger 
than most of the predictions,
but not too much considering the error range.



\subsection{Correlation with the rise rate of the polar field and the prediction of Cycle 25}
\label{sec:pol_rr}

\begin{figure}
\centering
\includegraphics[scale=0.3]{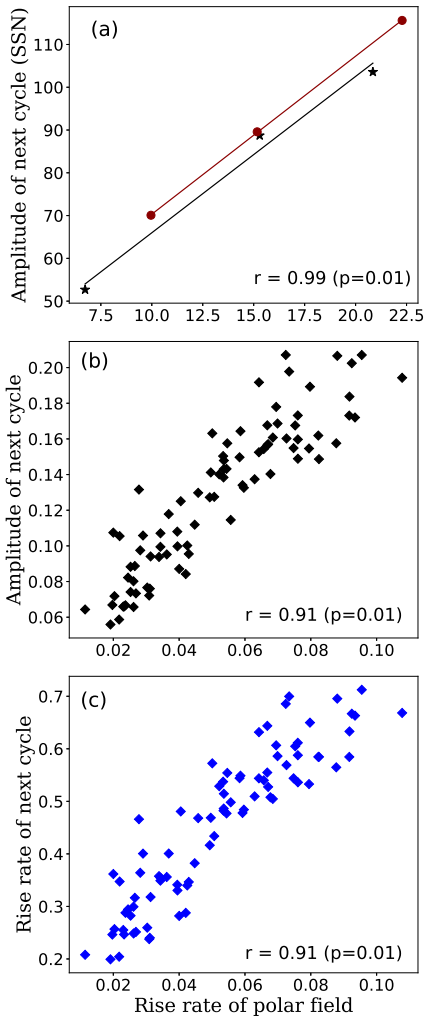}
\caption{(a) Scatter plot between rise rate of observed polar field and the amplitude of the next cycle SSN. Regression lines: $y=mx+c$, where m = $3.677\pm 0.876$  and c = $29.093\pm 13.517$ for the northern (asterisks) and m = $3.689\pm 0.912$ and c = $33.405\pm 15.311$ for southern hemispheres (filled circles).
(b) Same as (a) but from the dynamo model.
(c) Same as (b) but for the rise rate of the next sunspot cycle.
}
\label{fig:polr}
\end{figure}

Finally, further going one step backwards in the evolution of solar cycles, we find a very interesting relation that the rise rate of the polar field build-up (after reversal) has a correlation with the amplitude of the next cycle; see \Fig{fig:polr}(a). 
We note that here we have used the hemispheric data for the correlation. 
We find a similar strong correlation  ($r = 0.99$, $p = 0.01$)
for both SSN and SSA data.
We could compute this correlation 
in proxies of polar field, namely the $A(t)$ index and polar faculae count.
However, the timing of the polarity reversal is not determined in these data. Importantly, these data are very noisy 
and computing the rise rate in these data sometimes leads to poor correlation; see Table~2 of \citet{pawan21}.
We note that we had computed the average rise rate during the first three years after the polar field reversal. 
If we go beyond 3 years, then the polar field tends to saturate and the rise rate poorly correlates with the amplitude of the next cycle. 
Unfortunately, again the reliability of this relation cannot be proven based on only three data points. However, we find a strong relation between these two quantities in the dynamo model 
(again from Run 2DR2); see \Fig{fig:polr}(b).
We note that this relation is also strongly reproduced in other dynamo models; see Table~4 of \citet{pawan21}). 
As this relation holds good, we obviously expect a strong correlation between the rise rate of the polar field build-up and the rise rate of the next cycle, which is indeed seen in \Fig{fig:polr}(c).

The physics behind this correlation is not difficult to understand. In the \bl\ process, the decay and dispersal of tilted BMRs produces polar field in the Sun. When a sunspot cycle reaches its maximum, the polar field is usually reversed and then as the new BMRs emerge, the polar field increases 
(due to continuous supply of the trailing polarity flux from low latitudes) while the sunspot cycle decline. 
Hence, if the polar field in a cycle rises rapidly, then the toroidal field for the next cycle will also be amplified rapidly. This causes the next sunspot cycle to rise fast and also makes it strong.

One follow-up question is why the rise rate of the polar field build-up is not the same for all cycles. It is because the generation of poloidal field involves some randomness, particularly due to scatter in the BMR tilts \citep{JCS14, HCM17, KM17, Jha20} and the latitudinal positions of BMRs \citep{MKB17, KM18, Kar20}. In fact, there is indication that the decline phase of the cycle (during which the polar field is built up after reversal) is more irregular having many anti-hale and non-Joy BMRs \citep{Zhukova22, Mord22}, that can disturb the growth of the polar field considerably. 
Temporal variation in the meridional flow can also lead to a change in the polar field build-up \citep{Kar10}.  Due to these inherent randomnesses in the \bl\ process, even if two cycles decay identically, their corresponding polar field build up can be different.

In conclusion, if the correlation between the rise rate and the amplitude of the next cycle, as seen in the observed data (\Fig{fig:polr}(a)) and in the dynamo model (\Fig{fig:polr}(b)) really holds good in the Sun, then we can make prediction of the solar cycle a few years before the time of the previous polar field peak or the solar minimum. This considerably increases the temporal scope of the predictability of the solar cycle.  
Using the observed regression relation between the polar field rise rate and the amplitude of the next solar cycle (\Fig{fig:polr}(a)), we find the peak of the ongoing Cycle 25 
to be $137\pm 23$. 
Instead of hemispheric SSN data, if we use SSA data, and then convert the predicted value into SSN (using the regression relation between the SSN and SSA), we get the peak value to be $144 \pm 3$. 
So we clearly see that these two values are quite close to the one that 
we have obtained using WE2 relation ($138\pm26$) in \Sec{sec:pred_we2}. 

We note that in \citet{pawan21}, our earlier prediction for Cycle~25 was $120\pm25$, which is 
lower than the current prediction. 
 This is because there we have used the polar field value at 4 years after reversal when the field tends to saturate. In contrast, in the present work, we have used the average rise rate from the first three year's data. 
 Furthermore, in previous work, the regression relation based on the polar field data at 4 years of the reversal was not very tight.

\begin{table}
\centering
\caption{Comparison of our predictions for Solar Cycle 25 (P1: using the rise rate of the SSN, P2: using the rise rate of the previous cycle's polar field) with predictions by other groups who used observed polar precursor.}
\begin{tabular}{lllllcl} 
\cline{1-5}
Authors &&  Predicted SSN  && Time  \\ 
\cline{1-5}
 This work: P1   &&   $138\pm 26$     &&  $2024.5\pm 0.8$         \\
 ~~~~~~~~~~~~~~~~: P2 && $137\pm 23$          &&                              \\
 \cline{1-5}
 \citet{pawan21}   &&  $120\pm 25 $     &&  $---$         \\
 \cline{1-5}
\citet{wg}   &&  $126$  &&  $---$    \\
\cline{1-5}
\citet{HC19}   &&  $140.5\pm 2.5$  &&  $---$    \\
\cline{1-5}
 \citet{Pesnell18} &&  $135\pm 25$   && $2025.2\pm 1.5$    \\
\cline{1-5}
\citet{Petrovay18} &&    130   &&   Late 2024              \\
\cline{1-5}
\citet{Gopalswamy18} &&   148            &&   $---$         \\
\cline{1-5}
\citet{Bhowmik+Nandy}  && 118   && $2024\pm 1$ \\
\cline{1-5}
\citet{Jiang_2018} && $125\pm 32$ && $---$ \\
\cline{1-5}
\citet{UH18}   && 110      && $---$     \\
\cline{1-5}

\end{tabular}
\label{table2}
\end{table}

\section{Conclusions}
We have utilised a robust feature of \we\, namely, the rise rate of a cycle is strongly correlated with its amplitude (WE2)
and we have shown that a reliable 
prediction of a solar cycle can be made when the cycle has just past a few years from its minimum.
The ongoing solar cycle has passed about two years, and using this data, we predict that the amplitude of Cycle 25 will be 
$138\pm26$ and it will attain peak around mid to late 2024.
Hence, the ongoing cycle will be slightly stronger than the previous Cycle 24.
Our predicted strength is also significantly
larger than the NOAA/NASA Prediction Panel \footnote{\url{https://www.swpc.noaa.gov/news/solar-cycle-25-forecast-update}}, which predicted the peak SSN of Cycle 25 to be $115\pm10$.

We have shown that this prediction method (WE2) is based on a strong physical ground. If a polar field of a cycle is strong, then the next cycle has to be strong and it will also rise fast. Hence, our prediction based on the rise rate should be comparable to the one based on the polar field of the previous cycle; see \Tab{table2}. But this is not the
complete story.
If the polar field builds up (after its reversal)  rapidly, then the next cycle will be strong and vice versa. 
Therefore, we find a strong correlation between the rise rate of the polar field and the amplitude of the next cycle, both in observations and in dynamo models \citep[also see Appendix of][]{pawan21}.
Based on the observed regression relation between the rise rate 
of the previous cycle's polar field with the amplitude of the sunspot cycle, we predict the amplitude of the ongoing Cycle 25 to be 
$137\pm 23$.

Hence, our predictions made from both methods, namely using the rise rate of sunspot cycle and the rise rate of the previous cycle's polar field, match quite well. This agreement is because of the fact that they are linked. The polar field of the previous cycle gives rise to the toroidal field and the sunspot of the current cycle and thus, how rapidly the Sun builds up its polar field determines the rise rate of the next cycle. This link between the rate of build-up of the polar field with the amplitude of the next cycle suggest that we can predict the amplitude of the next solar cycle just after about 2-3 years of the reversal of its polar field (or up to about 9 years before the peak of a cycle).
Earlier \citet{pawan21} have shown that the prediction of the cycle can be made just after 4 years of the reversal of the polar field. 
Hence, this study, along with our present work, extends the scope of solar cycle prediction by a considerable amount of time.

\section*{Acknowledgements}
The authors thank the anonymous referee for a suggestion that made the title more accurate.
The authors also acknowledge financial support provided by ISRO/RESPOND (project No.
ISRO/RES/2/430/19-20).
\section*{Data Availability}
\label{sec:data}
We have used the SSN data available at SILSO\footnote{\url{http://sidc.oma.be/silso/DATA/SN_ms_tot_V2.0.txt}} and SSA from the Royal Greenwich Observatory (RGO)\footnote{\url{https://solarscience.msfc.nasa.gov/greenwch.shtml}}. The hemispheric SSN data has been collected from \citet{hem21}\footnote{\url{https://wwwbis.sidc.be/silso/extheminum}}.
Polar field data is taken from Wilcox Solar Observatory (WSO)\footnote{\url{http://wso.stanford.edu/Polar.html}}. 
Data from our dynamo models and the analyses codes can be shared upon a reasonable request. 



\bibliographystyle{mnras}
\bibliography{paper} 

\begin{thebibliography}{}
\makeatletter
\relax
\def\mn@urlcharsother{\let\do\@makeother \do\$\do\&\do\#\do\^\do\_\do\%\do\~}
\def\mn@doi{\begingroup\mn@urlcharsother \@ifnextchar [ {\mn@doi@}
  {\mn@doi@[]}}
\def\mn@doi@[#1]#2{\def\@tempa{#1}\ifx\@tempa\@empty \href
  {http://dx.doi.org/#2} {doi:#2}\else \href {http://dx.doi.org/#2} {#1}\fi
  \endgroup}
\def\mn@eprint#1#2{\mn@eprint@#1:#2::\@nil}
\def\mn@eprint@arXiv#1{\href {http://arxiv.org/abs/#1} {{\tt arXiv:#1}}}
\def\mn@eprint@dblp#1{\href {http://dblp.uni-trier.de/rec/bibtex/#1.xml}
  {dblp:#1}}
\def\mn@eprint@#1:#2:#3:#4\@nil{\def\@tempa {#1}\def\@tempb {#2}\def\@tempc
  {#3}\ifx \@tempc \@empty \let \@tempc \@tempb \let \@tempb \@tempa \fi \ifx
  \@tempb \@empty \def\@tempb {arXiv}\fi \@ifundefined
  {mn@eprint@\@tempb}{\@tempb:\@tempc}{\expandafter \expandafter \csname
  mn@eprint@\@tempb\endcsname \expandafter{\@tempc}}}

\bibitem[\protect\citeauthoryear{{Bertello}, {Pevtsov}, {Petrie}  \&
  {Keys}}{{Bertello} et~al.}{2014}]{Bertello14}
{Bertello} L.,  {Pevtsov} A.~A.,  {Petrie} G.~J.~D.,   {Keys} D.,  2014,
  \mn@doi [\solphys] {10.1007/s11207-014-0480-3}, \href
  {https://ui.adsabs.harvard.edu/abs/2014SoPh..289.2419B} {289, 2419}

\bibitem[\protect\citeauthoryear{{Bhowmik} \& {Nandy}}{{Bhowmik} \&
  {Nandy}}{2018}]{Bhowmik+Nandy}
{Bhowmik} P.,  {Nandy} D.,  2018, \mn@doi [Nature Communications]
  {10.1038/s41467-018-07690-0}, \href
  {https://ui.adsabs.harvard.edu/abs/2018NatCo...9.5209B} {9, 5209}

\bibitem[\protect\citeauthoryear{{Cameron} \& {Sch{\"u}ssler}}{{Cameron} \&
  {Sch{\"u}ssler}}{2007}]{CS07}
{Cameron} R.,  {Sch{\"u}ssler} M.,  2007, \mn@doi [\apj] {10.1086/512049},
  \href {http://adsabs.harvard.edu/abs/2007ApJ...659..801C} {659, 801}

\bibitem[\protect\citeauthoryear{{Cameron} \& {Sch{\"u}ssler}}{{Cameron} \&
  {Sch{\"u}ssler}}{2008}]{CS08}
{Cameron} R.,  {Sch{\"u}ssler} M.,  2008, \mn@doi [\apj] {10.1086/591079},
  \href {http://adsabs.harvard.edu/abs/2008ApJ...685.1291C} {685, 1291}

\bibitem[\protect\citeauthoryear{{Chatterjee}, {Nandy}  \&
  {Choudhuri}}{{Chatterjee} et~al.}{2004}]{CNC04}
{Chatterjee} P.,  {Nandy} D.,   {Choudhuri} A.~R.,  2004, \mn@doi [\aap]
  {10.1051/0004-6361:20041199}, \href
  {http://adsabs.harvard.edu/abs/2004A%26A...427.1019C} {427, 1019}

\bibitem[\protect\citeauthoryear{{Choudhuri}, {Chatterjee}  \&
  {Jiang}}{{Choudhuri} et~al.}{2007}]{CCJ07}
{Choudhuri} A.~R.,  {Chatterjee} P.,   {Jiang} J.,  2007, \mn@doi [Physical
  Review Letters] {10.1103/PhysRevLett.98.131103}, \href
  {http://adsabs.harvard.edu/abs/2007PhRvL..98m1103C} {98, 131103}

\bibitem[\protect\citeauthoryear{{Dikpati}, {Gilman}  \& {de Toma}}{{Dikpati}
  et~al.}{2008}]{Dik08}
{Dikpati} M.,  {Gilman} P.~A.,   {de Toma} G.,  2008, \mn@doi [\apjl]
  {10.1086/527360}, \href {http://adsabs.harvard.edu/abs/2008ApJ...673L..99D}
  {673, L99}

\bibitem[\protect\citeauthoryear{{Garg}, {Karak}, {Egeland}, {Soon}  \&
  {Baliunas}}{{Garg} et~al.}{2019}]{garg19}
{Garg} S.,  {Karak} B.~B.,  {Egeland} R.,  {Soon} W.,   {Baliunas} S.,  2019,
  \mn@doi [\apj] {10.3847/1538-4357/ab4a17}, \href
  {https://ui.adsabs.harvard.edu/abs/2019ApJ...886..132G} {886, 132}

\bibitem[\protect\citeauthoryear{{Gopalswamy}, {M{\"a}kel{\"a}}, {Yashiro}  \&
  {Akiyama}}{{Gopalswamy} et~al.}{2018}]{Gopalswamy18}
{Gopalswamy} N.,  {M{\"a}kel{\"a}} P.,  {Yashiro} S.,   {Akiyama} S.,  2018,
  \mn@doi [Journal of Atmospheric and Solar-Terrestrial Physics]
  {10.1016/j.jastp.2018.04.005}, \href
  {https://ui.adsabs.harvard.edu/abs/2018JASTP.176...26G} {176, 26}

\bibitem[\protect\citeauthoryear{{Guo}, {Jiang}  \& {Wang}}{{Guo}
  et~al.}{2021}]{wg}
{Guo} W.,  {Jiang} J.,   {Wang} J.-X.,  2021, \mn@doi [\solphys]
  {10.1007/s11207-021-01878-2}, \href
  {https://ui.adsabs.harvard.edu/abs/2021SoPh..296..136G} {296, 136}

\bibitem[\protect\citeauthoryear{{Hathaway}, {Wilson}  \&
  {Reichmann}}{{Hathaway} et~al.}{2002}]{Hat02}
{Hathaway} D.~H.,  {Wilson} R.~M.,   {Reichmann} E.~J.,  2002, \mn@doi
  [\solphys] {10.1023/A:1022425402664}, \href
  {https://ui.adsabs.harvard.edu/abs/2002SoPh..211..357H} {211, 357}

\bibitem[\protect\citeauthoryear{{Hazra} \& {Choudhuri}}{{Hazra} \&
  {Choudhuri}}{2019}]{HC19}
{Hazra} G.,  {Choudhuri} A.~R.,  2019, \mn@doi [\apj]
  {10.3847/1538-4357/ab2718}, \href
  {https://ui.adsabs.harvard.edu/abs/2019ApJ...880..113H} {880, 113}

\bibitem[\protect\citeauthoryear{{Hazra}, {Karak}, {Banerjee}  \&
  {Choudhuri}}{{Hazra} et~al.}{2015}]{Haz15}
{Hazra} G.,  {Karak} B.~B.,  {Banerjee} D.,   {Choudhuri} A.~R.,  2015, \mn@doi
  [\solphys] {10.1007/s11207-015-0718-8}, \href
  {http://adsabs.harvard.edu/abs/2015SoPh..290.1851H} {290, 1851}

\bibitem[\protect\citeauthoryear{{Hazra}, {Choudhuri}  \& {Miesch}}{{Hazra}
  et~al.}{2017}]{HCM17}
{Hazra} G.,  {Choudhuri} A.~R.,   {Miesch} M.~S.,  2017, \mn@doi [\apj]
  {10.3847/1538-4357/835/1/39}, \href
  {http://adsabs.harvard.edu/abs/2017ApJ...835...39H} {835, 39}

\bibitem[\protect\citeauthoryear{{Jha}, {Karak}, {Mandal}  \& {Banerjee}}{{Jha}
  et~al.}{2020}]{Jha20}
{Jha} B.~K.,  {Karak} B.~B.,  {Mandal} S.,   {Banerjee} D.,  2020, \mn@doi
  [\apjl] {10.3847/2041-8213/ab665c}, \href
  {https://ui.adsabs.harvard.edu/abs/2020ApJ...889L..19J} {889, L19}

\bibitem[\protect\citeauthoryear{{Jiang}, {Chatterjee}  \& {Choudhuri}}{{Jiang}
  et~al.}{2007}]{JCC07}
{Jiang} J.,  {Chatterjee} P.,   {Choudhuri} A.~R.,  2007, \mn@doi [\mnras]
  {10.1111/j.1365-2966.2007.12267.x}, \href
  {http://adsabs.harvard.edu/abs/2007MNRAS.381.1527J} {381, 1527}

\bibitem[\protect\citeauthoryear{{Jiang}, {Cameron}  \&
  {Sch{\"u}ssler}}{{Jiang} et~al.}{2014}]{JCS14}
{Jiang} J.,  {Cameron} R.~H.,   {Sch{\"u}ssler} M.,  2014, \mn@doi [\apj]
  {10.1088/0004-637X/791/1/5}, \href
  {http://adsabs.harvard.edu/abs/2014ApJ...791....5J} {791, 5}

\bibitem[\protect\citeauthoryear{Jiang, Wang, Jiao  \& Cao}{Jiang
  et~al.}{2018}]{Jiang_2018}
Jiang J.,  Wang J.-X.,  Jiao Q.-R.,   Cao J.-B.,  2018, \mn@doi [The
  Astrophysical Journal] {10.3847/1538-4357/aad197}, 863, 159

\bibitem[\protect\citeauthoryear{{Kane}}{{Kane}}{2010}]{Kane10}
{Kane} R.~P.,  2010, \mn@doi [\solphys] {10.1007/s11207-009-9466-y}, \href
  {http://adsabs.harvard.edu/abs/2010SoPh..261..209K} {261, 209}

\bibitem[\protect\citeauthoryear{{Karak}}{{Karak}}{2010}]{Kar10}
{Karak} B.~B.,  2010, \mn@doi [\apj] {10.1088/0004-637X/724/2/1021}, \href
  {http://adsabs.harvard.edu/abs/2010ApJ...724.1021K} {724, 1021}

\bibitem[\protect\citeauthoryear{{Karak}}{{Karak}}{2020}]{Kar20}
{Karak} B.~B.,  2020, \mn@doi [\apjl] {10.3847/2041-8213/abb93f}, \href
  {https://ui.adsabs.harvard.edu/abs/2020ApJ...901L..35K} {901, L35}

\bibitem[\protect\citeauthoryear{{Karak} \& {Choudhuri}}{{Karak} \&
  {Choudhuri}}{2011}]{KC11}
{Karak} B.~B.,  {Choudhuri} A.~R.,  2011, \mn@doi [\mnras]
  {10.1111/j.1365-2966.2010.17531.x}, \href
  {http://adsabs.harvard.edu/abs/2011MNRAS.410.1503K} {410, 1503}

\bibitem[\protect\citeauthoryear{{Karak} \& {Miesch}}{{Karak} \&
  {Miesch}}{2017}]{KM17}
{Karak} B.~B.,  {Miesch} M.,  2017, \mn@doi [\apj] {10.3847/1538-4357/aa8636},
  \href {http://adsabs.harvard.edu/abs/2017ApJ...847...69K} {847, 69}

\bibitem[\protect\citeauthoryear{{Karak} \& {Miesch}}{{Karak} \&
  {Miesch}}{2018}]{KM18}
{Karak} B.~B.,  {Miesch} M.,  2018, \mn@doi [\apjl] {10.3847/2041-8213/aaca97},
  \href {http://adsabs.harvard.edu/abs/2018ApJ...860L..26K} {860, L26}

\bibitem[\protect\citeauthoryear{{Karak}, {Mandal}  \& {Banerjee}}{{Karak}
  et~al.}{2018}]{KMB18}
{Karak} B.~B.,  {Mandal} S.,   {Banerjee} D.,  2018, \mn@doi [\apj]
  {10.3847/1538-4357/aada0d}, \href
  {https://ui.adsabs.harvard.edu/abs/2018ApJ...866...17K} {866, 17}

\bibitem[\protect\citeauthoryear{{Kitchatinov} \& {Olemskoy}}{{Kitchatinov} \&
  {Olemskoy}}{2011}]{KO11}
{Kitchatinov} L.~L.,  {Olemskoy} S.~V.,  2011, \mn@doi [Astronomy Letters]
  {10.1134/S0320010811080031}, \href
  {http://adsabs.harvard.edu/abs/2011AstL...37..656K} {37, 656}

\bibitem[\protect\citeauthoryear{{Kumar}, {Nagy}, {Lemerle}, {Karak}  \&
  {Petrovay}}{{Kumar} et~al.}{2021a}]{pawan21}
{Kumar} P.,  {Nagy} M.,  {Lemerle} A.,  {Karak} B.~B.,   {Petrovay} K.,  2021a,
  \mn@doi [\apj] {10.3847/1538-4357/abdbb4}, \href
  {https://ui.adsabs.harvard.edu/abs/2021ApJ...909...87K} {909, 87}

\bibitem[\protect\citeauthoryear{{Kumar}, {Karak}  \& {Vashishth}}{{Kumar}
  et~al.}{2021b}]{KKV21}
{Kumar} P.,  {Karak} B.~B.,   {Vashishth} V.,  2021b, \mn@doi [\apj]
  {10.3847/1538-4357/abf0a1}, \href
  {https://ui.adsabs.harvard.edu/abs/2021arXiv210311754K} {913, 65}

\bibitem[\protect\citeauthoryear{{Makarov}, {Makarova}  \&
  {Sivaraman}}{{Makarov} et~al.}{1989}]{MMS89}
{Makarov} V.~I.,  {Makarova} V.~V.,   {Sivaraman} K.~R.,  1989, \mn@doi
  [\solphys] {10.1007/BF00146211}, \href
  {https://ui.adsabs.harvard.edu/abs/1989SoPh..119...45M} {119, 45}

\bibitem[\protect\citeauthoryear{{Mandal}, {Karak}  \& {Banerjee}}{{Mandal}
  et~al.}{2017}]{MKB17}
{Mandal} S.,  {Karak} B.~B.,   {Banerjee} D.,  2017, \mn@doi [\apj]
  {10.3847/1538-4357/aa97dc}, \href
  {http://adsabs.harvard.edu/abs/2017ApJ...851...70M} {851, 70}

\bibitem[\protect\citeauthoryear{{Mordvinov}, {Karak}, {Banerjee}, {Golubeva},
  {Khlystova}, {Zhukova}  \& {Kumar}}{{Mordvinov} et~al.}{2022}]{Mord22}
{Mordvinov} A.~V.,  {Karak} B.~B.,  {Banerjee} D.,  {Golubeva} E.~M.,
  {Khlystova} A.~I.,  {Zhukova} A.~V.,   {Kumar} P.,  2022, \mn@doi [\mnras]
  {10.1093/mnras/stab3528}, \href
  {https://ui.adsabs.harvard.edu/abs/2022MNRAS.510.1331M} {510, 1331}

\bibitem[\protect\citeauthoryear{{Mu{\~n}oz-Jaramillo}, {Dasi-Espuig},
  {Balmaceda}  \& {DeLuca}}{{Mu{\~n}oz-Jaramillo} et~al.}{2013}]{Muno13}
{Mu{\~n}oz-Jaramillo} A.,  {Dasi-Espuig} M.,  {Balmaceda} L.~A.,   {DeLuca}
  E.~E.,  2013, \mn@doi [\apjl] {10.1088/2041-8205/767/2/L25}, \href
  {http://adsabs.harvard.edu/abs/2013ApJ...767L..25M} {767, L25}

\bibitem[\protect\citeauthoryear{{Nandy}}{{Nandy}}{2021}]{Nandy21}
{Nandy} D.,  2021, \mn@doi [\solphys] {10.1007/s11207-021-01797-2}, \href
  {https://ui.adsabs.harvard.edu/abs/2021SoPh..296...54N} {296, 54}

\bibitem[\protect\citeauthoryear{{Pesnell} \& {Schatten}}{{Pesnell} \&
  {Schatten}}{2018}]{Pesnell18}
{Pesnell} W.~D.,  {Schatten} K.~H.,  2018, \mn@doi [\solphys]
  {10.1007/s11207-018-1330-5}, \href
  {https://ui.adsabs.harvard.edu/abs/2018SoPh..293..112P} {293, 112}

\bibitem[\protect\citeauthoryear{{Petrovay}}{{Petrovay}}{2020}]{Petrovay20}
{Petrovay} K.,  2020, \mn@doi [Living Reviews in Solar Physics]
  {10.1007/s41116-020-0022-z}, \href
  {https://ui.adsabs.harvard.edu/abs/2020LRSP...17....2P} {17, 2}

\bibitem[\protect\citeauthoryear{{Petrovay}, {Nagy}, {Gerj{\'a}k}  \&
  {Juh{\'a}sz}}{{Petrovay} et~al.}{2018}]{Petrovay18}
{Petrovay} K.,  {Nagy} M.,  {Gerj{\'a}k} T.,   {Juh{\'a}sz} L.,  2018, \mn@doi
  [Journal of Atmospheric and Solar-Terrestrial Physics]
  {10.1016/j.jastp.2017.12.011}, \href
  {https://ui.adsabs.harvard.edu/abs/2018JASTP.176...15P} {176, 15}

\bibitem[\protect\citeauthoryear{{Pipin} \& {Kosovichev}}{{Pipin} \&
  {Kosovichev}}{2011}]{pipin11}
{Pipin} V.~V.,  {Kosovichev} A.~G.,  2011, \mn@doi [\apj]
  {10.1088/0004-637X/741/1/1}, \href
  {https://ui.adsabs.harvard.edu/abs/2011ApJ...741....1P} {741, 1}

\bibitem[\protect\citeauthoryear{{Pipin} \& {Sokoloff}}{{Pipin} \&
  {Sokoloff}}{2011}]{pipin11b}
{Pipin} V.~V.,  {Sokoloff} D.~D.,  2011, \mn@doi [\physscr]
  {10.1088/0031-8949/84/06/065903}, \href
  {https://ui.adsabs.harvard.edu/abs/2011PhyS...84f5903P} {84, 065903}

\bibitem[\protect\citeauthoryear{{Priyal}, {Banerjee}, {Karak},
  {Mu{\~n}oz-Jaramillo}, {Ravindra}, {Choudhuri}  \& {Singh}}{{Priyal}
  et~al.}{2014}]{Priy14}
{Priyal} M.,  {Banerjee} D.,  {Karak} B.~B.,  {Mu{\~n}oz-Jaramillo} A.,
  {Ravindra} B.,  {Choudhuri} A.~R.,   {Singh} J.,  2014, \mn@doi [\apjl]
  {10.1088/2041-8205/793/1/L4}, \href
  {http://adsabs.harvard.edu/abs/2014ApJ...793L...4P} {793, L4}

\bibitem[\protect\citeauthoryear{{Schatten}, {Scherrer}, {Svalgaard}  \&
  {Wilcox}}{{Schatten} et~al.}{1978}]{Sch78}
{Schatten} K.~H.,  {Scherrer} P.~H.,  {Svalgaard} L.,   {Wilcox} J.~M.,  1978,
  \mn@doi [\grl] {10.1029/GL005i005p00411}, \href
  {http://adsabs.harvard.edu/abs/1978GeoRL...5..411S} {5, 411}

\bibitem[\protect\citeauthoryear{{Upton} \& {Hathaway}}{{Upton} \&
  {Hathaway}}{2018}]{UH18}
{Upton} L.~A.,  {Hathaway} D.~H.,  2018, \mn@doi [\grl] {10.1029/2018GL078387},
  \href {https://ui.adsabs.harvard.edu/abs/2018GeoRL..45.8091U} {45, 8091}

\bibitem[\protect\citeauthoryear{{Veronig}, {Jain}, {Podladchikova},
  {P{\"o}tzi}  \& {Clette}}{{Veronig} et~al.}{2021}]{hem21}
{Veronig} A.~M.,  {Jain} S.,  {Podladchikova} T.,  {P{\"o}tzi} W.,   {Clette}
  F.,  2021, \mn@doi [\aap] {10.1051/0004-6361/202141195}, \href
  {https://ui.adsabs.harvard.edu/abs/2021A&A...652A..56V} {652, A56}

\bibitem[\protect\citeauthoryear{{Waldmeier}}{{Waldmeier}}{1935}]{W35}
{Waldmeier} M.,  1935, Astronomische Mitteilungen der Eidgen{\"o}ssischen
  Sternwarte Zurich, \href {http://adsabs.harvard.edu/abs/1935MiZur..14..105W}
  {14, 105}

\bibitem[\protect\citeauthoryear{{Wang} \& {Sheeley}}{{Wang} \&
  {Sheeley}}{2009}]{WS09}
{Wang} Y.-M.,  {Sheeley} N.~R.,  2009, \mn@doi [\apjl]
  {10.1088/0004-637X/694/1/L11}, \href
  {http://adsabs.harvard.edu/abs/2009ApJ...694L..11W} {694, L11}

\bibitem[\protect\citeauthoryear{{Zhukova}, {Khlystova}, {Abramenko}  \&
  {Sokoloff}}{{Zhukova} et~al.}{2022}]{Zhukova22}
{Zhukova} A.,  {Khlystova} A.,  {Abramenko} V.,   {Sokoloff} D.,  2022, \mn@doi
  [\mnras] {10.1093/mnras/stac597}, \href
  {https://ui.adsabs.harvard.edu/abs/2022MNRAS.tmp..593Z} {}

\makeatother
\end{thebibliography}




\bsp	
\label{lastpage}
\end{document}